\documentclass[final, 1p, 12pt, times]{elsarticle}

\usepackage{amssymb}
\usepackage{amsmath}

\usepackage{xcolor} 
\usepackage[braket,qm]{qcircuit}
\usepackage{qcircuit}

\usepackage{xcolor}
\usepackage{subcaption} 
\usepackage{float}
\usepackage{booktabs}
\usepackage{multirow}

\definecolor{mediumseagreen}{HTML}{3CB371}
\newcommand{\X}{\mathbf{X}}
\newcommand{\Y}{\mathbf{Y}}

\journal{Journal of Computational and Applied Mathematics}

\begin{document}

\begin{frontmatter}



\title{Quantum Neural Network Architectures for Multivariate Time-Series Forecasting} 



\author[1]{S. Ranilla-Cortina}
\author[2]{D. A. Aranda}
\author[3]{J. Ballesteros}
\author[2]{J. Bonilla}
\author[2]{N. Monrio}
\author[1]{Elías F. Combarro}
\author[1]{J. Ranilla}

\affiliation[1]{
organization={Computer Science Department},
addressline={University of Oviedo},
Country={Spain}}

\affiliation[2]{
organization={IA Orbital},
addressline={ARQUIMEA Research Center},
Country={Spain}}

\affiliation[3]{
organization={Instituto Tecnológico y de Energías Renovables, S.A.},
Country={Spain}}

\begin{abstract}

In this paper, we address the challenge of multivariate time-series forecasting using quantum machine learning techniques. We introduce adaptation strategies that extend variational quantum circuit models, traditionally limited to univariate data, toward the multivariate setting, exploring both purely quantum and hybrid quantum-classical formulations. First, we extend and benchmark several VQC-based and hybrid architectures to systematically evaluate their capacity to model cross-variable dependencies. Second, building upon these foundations, we introduce the iQTransformer, a novel quantum transformer architecture that integrates a quantum self-attention mechanism within the iTransformer framework, enabling a quantum-native representation of inter-variable relationships. Third, we provide a comprehensive empirical evaluation on both synthetic and real-world datasets, showing that quantum-based models may achieve competitive or superior forecasting accuracy with fewer trainable parameters and faster convergence than state-of-the-art classical and quantum baselines in some cases. These contributions highlight the potential of quantum-enhanced architectures as efficient and scalable tools for advancing multivariate time-series forecasting.

\end{abstract}




\end{frontmatter}




\section{Introduction}\label{sec:intro}

Quantum Machine Learning (QML) is an emerging interdisciplinary field at the intersection of machine learning and quantum computing~\cite{schuld2021machine}, aiming to exploit principles such as superposition, entanglement, and interference to improve data processing and pattern recognition. Recent advances have demonstrated its applicability to diverse tasks, including classification, reinforcement learning, and generative modeling~\cite{schuld2021machine, biamonte2017quantum, bergholm2018pennylane}. More recently, preliminary studies have begun to explore its application to \textit{time-series forecasting}, showing promising results but still facing important challenges when moving beyond univariate settings~\cite{fellner2025quantum}.

Time-series forecasting is critical in domains such as finance, energy systems, healthcare, and climate science, where accurate predictions of multiple interdependent variables guide decision-making~\cite{zhou2021informer, liu2023itransformer, lim2021survey}. Traditional machine learning and deep learning models, including Recurrent Neural Networks (RNNs), Long Short-Term Memory (LSTM) networks, and Transformer-based architectures, have achieved remarkable success in these tasks~\cite{hochreiter1997lstm, vaswani2017attention}. However, they often require large datasets, incur high computational costs, and may struggle to efficiently capture complex inter-variable dependencies as the number of channels grows~\cite{liu2023itransformer, nonstationarytransformerse}. These limitations motivate the exploration of quantum-enhanced models that promise more compact parameterizations, faster convergence, and potentially new representational advantages.

Most existing QML approaches for forecasting have been restricted to the \textit{univariate case}~\cite{fellner2025quantum, rivera20231d}, which does not reflect the multivariate structure of real-world applications where cross-variable dependencies are essential. Extending QML to the \textit{multivariate setting} introduces additional challenges: (i) efficiently encoding correlated variables into quantum states, (ii) designing circuits capable of modeling cross-variable dependencies, and (iii) balancing expressivity with the hardware limitations of near-term devices. Addressing these challenges is key to evaluating the practical value of quantum forecasting models.

In this work, we take a step in this direction by benchmarking several QML architectures for multivariate time-series forecasting. We contrast \textit{independent-channel designs}, where each variable is modeled by a separate quantum circuit, with \textit{joint-circuit approaches} that encode multiple variables simultaneously. In addition, we investigate \textit{hybrid quantum-classical models} that combine variational quantum circuits (VQCs) with classical layers, such as multilayer perceptrons (MLPs) or encoder-decoder pipelines, to enhance scalability and flexibility. Beyond these baselines, we introduce the \textit{iQTransformer}, a novel hybrid architecture that integrates a quantum self-attention mechanism into the recently proposed iTransformer framework. This design combines the inverted tokenization strategy of the iTransformer with quantum self-attention, providing a quantum-native approach to capture inter-variable dependencies in multivariate forecasting tasks, thereby overcoming a key limitation of prior QML approaches. We perform experiments on both a synthetic three-channel Lorenz dataset and a real-world seven-channel operational energy dataset. Our findings show that quantum-based models may achieve performance comparable to or exceeding that of classical baselines, while requiring fewer training steps. These highlight the potential of quantum-enhanced architectures as a promising direction to advance efficient and scalable forecasting in complex multivariate scenarios.

The remainder of this paper is structured as follows. Section~\ref{sec:related} reviews related work in quantum and classical time-series forecasting. Section~\ref{sec:approaches} introduces QML architectures adapted for the multivariate setting. Section~\ref{sec:base} presents the proposed iQTransformer model, while Section~\ref{sec:setup} describes the experimental setup. Section~\ref{sec:results} reports results and analysis, and Section~\ref{sec:conclusions} concludes with directions for future work.

\section{Related Work} \label{sec:related}

QML has rapidly evolved from proof-of-concept algorithms into increasingly sophisticated hybrid pipelines that combine parametrized quantum circuits with classical optimizers~\cite{schuld2021machine}. In time-series prediction, recent comparative studies have benchmarked variational quantum models against classical baselines, highlighting both scenarios of potential quantum advantage and current hardware limitations~\cite{fellner2025quantum}.

An important contribution is the \textit{Quantum Convolutional Neural Network} (QCNN), first proposed by Cong et al.~\cite{cong2019quantum} and later adapted to data classification by Hur et al.~\cite{hur2022quantum}. QCNNs alternate trainable convolutional filters with pooling layers, typically implemented via partial measurements and controlled operations that progressively reduce the number of active qubits while preserving entanglement. This design enables hierarchical feature extraction in the quantum domain and has served as a structured baseline for sequence learners, including early attempts at univariate forecasting and classification~\cite{rivera20231d}.

Another key line of work is based on \textit{data re-uploading} techniques~\cite{perez2020data}, which enhance expressivity by repeatedly encoding classical features across circuit layers, effectively creating a universal quantum classifier. These methods have been adapted to sequence modeling, enabling quantum models to process temporal data with greater flexibility, although most applications remain restricted to single-channel time series and the multivariate case is still underexplored.

On the classical side, deep learning has achieved remarkable progress in time-series forecasting, particularly with Transformer-based models. Architectures such as the \textit{Informer}~\cite{zhou2021informer} improve efficiency on long sequences, while the more recent \textit{iTransformer}~\cite{liu2023itransformer} redefined tokenization by treating input variables as tokens, thus enabling direct modeling of cross-variable dependencies. Variants like the \textit{Non-stationary Transformer}~\cite{nonstationarytransformerse} further address distributional shifts and long-horizon dependencies.

Despite these advances, QML for \textit{multivariate} time-series forecasting remains limited. Existing quantum approaches often rely on independent per-channel models, which fail to capture inter-variable correlations~\cite{fellner2025quantum, chen2024quantum}. Hybrid quantum-classical designs provide a promising direction: quantum circuits can encode complex temporal and cross-variable patterns, while classical layers handle aggregation, scaling, and long-horizon prediction.

In this context, our work's contribution is two-fold: (i) extending and benchmarking several state-of-the-art QML architectures, including independent-channel VQCs, joint-circuit designs, and hybrid encoder-decoder pipelines to the multivariate setting and (ii) introducing the \textit{iQTransformer}, which integrates a Quantum Self-Attention Neural Network (QSANN)~\cite{Li2024QSANN} into the iTransformer backbone, thereby providing a quantum-native mechanism to capture inter-variable dependencies. This combination addresses a key gap in current literature by uniting advances in classical forecasting with the representational potential of quantum models.

\section{Adapting Quantum Models for Multivariate Time-Series Forecasting} 
\label{sec:approaches}

In multivariate time-series forecasting, we are given a historical sequence $\X = \left[x_1, \ldots, x_T\right]^\top \in \mathbb{R}^{T \times C}$, where $T$ is the lookback window length and $C$ denotes the number of input channels or variates. Each observation $x_t \in \mathbb{R}^C$ contains the simultaneous measurements of all $C$ variates at time $t$. The forecasting task is to predict the next $S$ steps $\Y = \left[x_{T+1}, \ldots, x_{T+S}\right]^\top \in \mathbb{R}^{S \times C}$, where $S$ is the forecasting horizon. For convenience, we denote $\X_{t,:}$ as the simultaneously recorded time points at the step $t$, and $\X_{:,c}$ as the whole time series of each variate indexed by $c$. Depending on the application, one may predict all $C$ channels or focus on a specific target channel $c_{\text{tar}}$, whose trajectory $\Y_{:,c_{\text{tar}}} \in \mathbb{R}^S$ is estimated from the past history of all channels.

Most existing quantum machine learning approaches for time-series forecasting have been designed for the univariate case~\cite{fellner2025quantum, rivera20231d, chen2024quantum}, 
where $C=1$ and the models only learn from single-channel temporal dependencies. 
While this setting provides a simplified benchmark, it is far from the multivariate structure found in real-world applications, 
such as finance, energy, or climate systems, where cross-channel dependencies are essential. 
To move towards realistic use cases, we consider here the multivariate setting ($C > 1$) and investigate how to adapt VQCs and other state-of-the-art quantum machine learning architectures to this more challenging task.


In the remainder of this section, we present relatively simple adaptations of VQCs and hybrid quantum-classical models, extending architectures originally designed for univariate settings to the more realistic multivariate case. These models serve as baselines to explore the fundamental challenges of multivariate quantum forecasting. 

\begin{figure}
    \centering
    \includegraphics[width=1\linewidth]{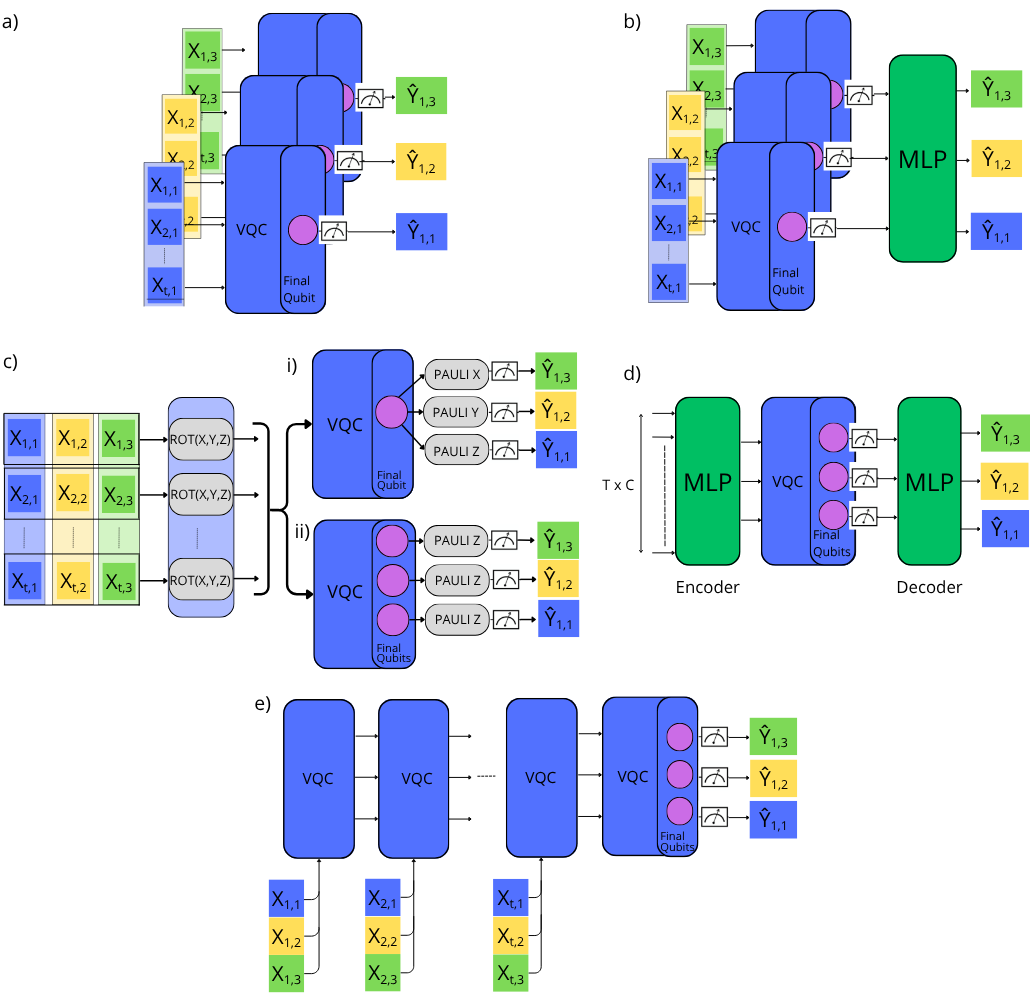}
    \caption{Different VQC-based architectures: 
(a) \textbf{Naive baseline:} independent VQC per channel, each modeled by the same univariate circuit without cross-channel interaction. 
(b) \textbf{Hybrid baseline:} independent VQCs per channel followed by a shallow MLP aggregating the quantum outputs to capture cross-channel correlations. 
(c) \textbf{Dense Embedding:} employs a dense rotational encoding embedding up to three input channels per qubit, capturing correlations directly at the quantum level. Two output variants are considered: (i) single-qubit with Pauli $\{X, Y, Z\}$ measurements, and (ii) three-qubit with Pauli-$Z$ measurements per channel. 
(d) \textbf{Encoder–decoder:} an encoder maps the input time series to the quantum space, a VQC processes the embedded features, and a decoder projects the outputs back to the forecasting space. 
(e) \textbf{Data re-uploading:} each time step is sequentially encoded through alternating encoding and variational layers.
 }
    \label{fig:multi_vqc}
\end{figure}

\subsection{Independent Channel VQC}
\label{subsec:indep_vqc}

As a naive baseline, we consider an independent quantum model applied separately to each channel. That is, a variational quantum circuit designed for univariate forecasting is replicated $C$ times, one for each channel, without any mechanism for modeling cross-channel dependencies. Each input series $\mathbf{X}_{:,c}$ is normalized prior to quantum feature encoding to ensure compatibility with the encoding scheme. This preprocessing step is common to all VQCs, as many embedding methods (e.g., angle or amplitude encoding) require input features to be scaled to a bounded interval such as $[0,1]$.

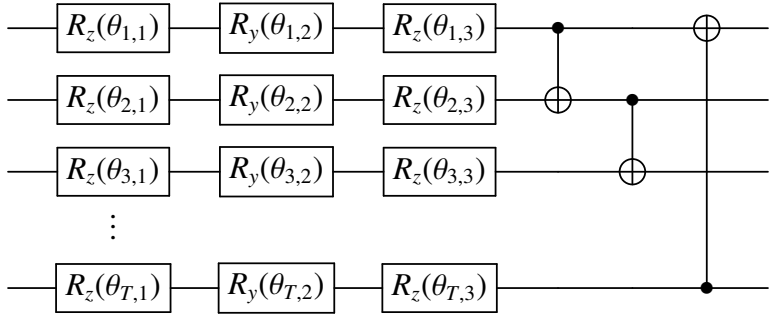
\begin{figure}[t]
\centering
\[
\Qcircuit @C=1.5em @R=0.7em {
    & \gate{R_z(\theta_{1,1})} & \gate{R_y(\theta_{1,2})} & \gate{R_z(\theta_{1,3})} & \ctrl{1} & \qw      & \targ     & \qw         \\
    & \gate{R_z(\theta_{2,1})} & \gate{R_y(\theta_{2,2})} & \gate{R_z(\theta_{2,3})} & \targ    & \ctrl{1} & \qw      & \qw      \\
    & \gate{R_z(\theta_{3,1})} & \gate{R_y(\theta_{3,2})} & \gate{R_z(\theta_{3,3})} & \qw      & \targ    & \qw & \qw      \\
    & \vdots                   &                         &                           &          &          &          &     \\
    &                          &                         &                           &          &          &          &     \\
    & \gate{R_z(\theta_{T,1})} & \gate{R_y(\theta_{T,2})} & \gate{R_z(\theta_{T,3})} & \qw      & \qw      & \ctrl{-5}& \qw }
\]
\caption{Variational quantum circuit used as a generic VQC ansatz. Each layer consists of single-qubit $R_y$ rotations followed by a ring of CNOT gates. $\theta$ correspond to trainable parameters. This block is repeated $p$ times to enhance expressivity.}
\label{fig:vqc_ansatz}
\end{figure}

In this baseline, each channel is modeled independently through a VQC performing single-step forecasting ($S=1$). The forecasting pipeline for one channel proceeds as follows. The classical input series $\mathbf{X}_{:,c} \in \mathbb{R}^T$ is encoded into a quantum state by an unitary $U_{\mathrm{enc}}$, acting on $T$ qubits initially prepared in the state $|0\rangle^{\otimes T}$. We employ angle encoding via $R_y$ rotations, such that each past observation $x_t$ (where $t=1,\dots,T$) is mapped to a qubit state as
\begin{equation}
|x_t\rangle = R_y(\pi x_t)\,|0\rangle \,.
\label{eq:encoding_ry}
\end{equation}
This choice of encoding is not intrinsic to the model and could be replaced by alternative schemes, such as amplitude or hybrid embeddings.

After encoding, a parameterized variational block $U_{\mathrm{var}}(\theta)$ is applied to introduce trainable correlations among qubits. 
The chosen ansatz is composed of layers combining single-qubit rotations and entangling CNOT gates, and the block is repeated $p$ times to increase expressivity. 
This generic circuit, illustrated in Fig.~\ref{fig:vqc_ansatz}, serves as our first reference implementation.

Finally, for each channel $c=1,\dots,C$ the expectation value of the Pauli-$Z$ operator is estimated on a designated qubit (here chosen as qubit~$0$) from repeated measurements in the computational basis:
\begin{equation}
Z_c = \langle 0^{\otimes n}| U_{\mathrm{enc}}^\dagger U_{\mathrm{var}}^\dagger(\theta)\, Z_0\, U_{\mathrm{var}}(\theta) U_{\mathrm{enc}} |0^{\otimes n}\rangle \,.
\end{equation}
Collecting these across all channels yields
\begin{equation}
\mathbf{Z} = \left[ Z_1, Z_2, \ldots, Z_C \right] \in [-1,1]^C \,,
\end{equation}
which forms the raw quantum output vector of the multi-channel system. $\mathbf{Z}$ is then rescaled to match the normalized data domain as $\hat{\mathbf{Y}} = (\mathbf{Z} + 1)/2$. A schematic overview of this baseline architecture is shown in Fig.~\ref{fig:multi_vqc}a. Notice this baseline is only compatible with single-step forecasting ($S=1$).

\subsection{VQC combined with MLP}
\label{subsec:vqc_mlp}

In this variant, we extend the independent channel VQC baseline by appending a shallow multilayer perceptron (MLP) $\phi_{\mathrm{mlp}}:\mathbb{R}^{C}\to\mathbb{R}^{C\times S}$ that aggregates the outputs of the quantum circuits across channels. 
This allows the model to capture cross-channel correlations through classical post-processing, yielding a hybrid quantum-classical design. 

As described in Subsec.~\ref{subsec:indep_vqc}, the quantum circuits output is a normalized vector $\mathbf{Z} \in [-1,1]^C$. Here, the post-processing MLP $\phi_{\mathrm{mlp}}$ computes the final prediction $\hat{\Y}$ via two affine layers with a ReLU activation in between as
\begin{equation}
\hat{\Y} = \phi_{\mathrm{mlp}}(\mathbf{Z}) = \mathbf{W}_2 \,\text{ReLU}(\mathbf{W}_1 \mathbf{Z} + b_1) + b_2 \,,
\end{equation}
where $\mathbf{W}_1,\mathbf{W}_2$ and $b_1,b_2$ are trainable parameters. 
The first layer maps $\mathbf{Z}$ to $\mathbb{R}^{2C\times S}$ and the second projects back to $\mathbb{R}^{C\times S}$, matching the forecasting horizon $S$ across all $C$ channels. In contrast to the purely quantum baseline of Subsec.~\ref{subsec:indep_vqc}, this hybrid architecture allows the model to learn dependencies between channels and supports forecasting horizons $S > 1$ simply by adjusting the output dimension of the final affine layer. A schematic of this hybrid baseline is shown in Fig.~\ref{fig:multi_vqc}b.

\subsection{Dense Embedding Variational Circuits}
\label{subsec:dense_vqc}

In this approach, instead of using the standard $R_y$ encoding from Eq.~\eqref{eq:encoding_ry} we employ a denser rotational encoding that allows multiple channels to be embedded within the same qubit: 
\begin{equation}
|\X_{t,1},\X_{t,2},\X_{t,3}\rangle 
= R_z(\pi \X_{t,3})\,R_y(\pi \X_{t,2})\,R_z(\pi \X_{t,1})\,|+\rangle \,,
\end{equation}
where $t \in \{1, 2, \dots, T\}$, and this allows up to three channels per qubit. Note that the first $R_z$ rotation would leave $|0\rangle$ invariant; therefore, in this case we initialize with the superposition state $|+\rangle = (|0\rangle + |1\rangle)/\sqrt{2}$ to enable full rotational expressivity. Unlike Subsec.~\ref{subsec:vqc_mlp}, correlations are handled directly at the quantum level instead of by classical post-processing.

For the output stage we considered two variants: 
(i) using a single qubit and assigning each channel to the expectation value of one Pauli operator from the set $\{X,Y,Z\}$, 
or (ii) measuring the Pauli-$Z$ operator on three different qubits, each corresponding to one channel. A schematic of these denser baselines is shown in Fig.~\ref{fig:multi_vqc}c. Although the case with three channels is presented here for illustration, the approach can be readily extended to a larger number of channels by employing denser quantum encodings (allowing multiple classical features to be mapped per qubit) and by measuring additional qubits and/or a broader set of quantum observables.

\subsection{Hybrid VQC with classical pre/post-processing}
\label{subsec:hybrid_vqc}

This baseline follows the generic VQC architecture shown in Fig.~\ref{fig:vqc_ansatz} but is framed within a hybrid design. A classical encoder $\phi_{\mathrm{enc}}:\mathbb{R}^{C\times T}\to\mathbb{R}^{n}$ maps the flattened input window into the quantum dimension, the quantum circuit generates a vector of measurements, and a classical decoder $\phi_{\mathrm{dec}}:\mathbb{R}^{n}\to\mathbb{R}^{C\times S}$ maps these features back to the forecasting space. In contrast to the independent-channel approach of Subsec.~\ref{subsec:indep_vqc}, the number of qubits, denoted by $n$, is a free hyperparameter, decoupled from the number of input channels $C$.

Formally, the encoder $\phi_{\mathrm{enc}}$ maps the flattened input $\X \in \mathbb{R}^{C\times T}$ into the quantum space through two trainable affine layers with a ReLU activation in between:
\begin{equation}
\phi_{\mathrm{enc}} (\X) = \mathbf{W}_2 \,\text{ReLU}(\mathbf{W}_1 \X + b_1) + b_2 \,,
\end{equation}
where $\mathbf{W}_1,\mathbf{W}_2$ and $b_1,b_2$ are trainable parameters. 
The first affine layer maps the input $\X \in \mathbb{R}^{C\times T}$ to an intermediate representation in $\mathbb{R}^{2n}$, and the second layer projects it back to $\mathbb{R}^n$. The resulting encoder output vector is then encoded on $n$ qubits via $R_y$ rotations and processed by the quantum circuit, producing a vector of Pauli-$Z$ expectation values,
\begin{equation}
\mathbf{Z} = \left[ \langle Z_1 \rangle, \ldots, \langle Z_n \rangle \right] \in [-1,1]^n \,,
\end{equation}
where $\langle Z_i \rangle$ corresponds to the measurement of the Pauli-$Z$ operator on qubit $i$.

The decoder $\phi_{\mathrm{dec}}$ is implemented analogously to the MLP in Subsec.~\ref{subsec:vqc_mlp}, 
with two affine layers and a ReLU activation in between:
\begin{equation}
\hat{\Y} = \phi_{\mathrm{dec}} (\mathbf{Z}) = \mathbf{W}_4 \,\text{ReLU}(\mathbf{W}_3 \mathbf{Z} + b_3) + b_4 \,,
\end{equation}
where $\mathbf{W}_3,\mathbf{W}_4$ and $b_3,b_4$ are trainable parameters.

This model differs from previous baselines in two ways: (i) the number of qubits $n$ is a free hyperparameter, and (ii) the circuit outputs a full feature vector $\mathbf{Z} \in \mathbb{R}^n$ by measuring all qubits. 
As in Subsec.~\ref{subsec:vqc_mlp}, forecasting horizons $S>1$ are naturally supported by adjusting the output dimension of the decoder. A schematic of this hybrid baseline is shown in Fig.~\ref{fig:multi_vqc}d.

This encoder–decoder formulation is general: the encoder maps classical inputs into any quantum dimension and the decoder projects outputs back to the forecasting space. This flexibility decouples the choice of quantum circuit size from the data dimensions, allowing the number of qubits $n$ to be freely selected as a modeling hyperparameter.

\subsection{Data Re-uploading Model}
\label{subsec:reuploading}

We also implemented a data re-uploading model similar to the one proposed in~\cite{fellner2025quantum}. 
Here the number of qubits equals the number of channels $C$, alternating encoding and variational layers. 
The total number of encoding layers is equal to the lookback window $T$, so that each temporal step is sequentially injected into the circuit before predicting the next value. 

Formally, at each time step $t \in \{1, 2, \dots, T\}$, the $t$-th encoding layer maps the input onto the quantum state via $R_y$ rotations,
\begin{equation}
|\psi^{(t)}\rangle = U_{\mathrm{var}}(\theta^{(t)}) \, U_{\mathrm{enc}}(\X_{t,:}) \,|\psi^{(t-1)}\rangle \,, 
\end{equation}
where $|\psi^{(0)}\rangle = |0\rangle^{\otimes C}$, $U_{\mathrm{enc}}$ denotes the feature encoding applied across the $C$ qubits, and $U_{\mathrm{var}}(\theta^{(t)})$ is the parameterized variational block at step $t$. 
After $T$ such re-uploading layers, the final quantum state is measured in the $Z$ basis across all qubits,
\begin{equation}
\mathbf{Z} = \left[\langle Z_1 \rangle, \ldots, \langle Z_C \rangle\right] \in [-1,1]^C \,,
\end{equation}
and rescaled to the interval $[0,1]$ to produce the final prediction
$\hat{\Y}$.

Without classical post-processing, this model supports only single-step forecasting ($S=1$).
A schematic of this model is in Fig.~\ref{fig:multi_vqc}e.

\section{Quantum Transformer model for multivariate forecasting} \label{sec:base}

This section introduces the architectural foundations of our quantum transformer model. We first revisit the classical iTransformer, which serves as the backbone for our design, and then describe the proposed quantum self-attention mechanism and its integration into the complete \textit{iQTransformer} pipeline.

The iTransformer~\cite{liu2023itransformer} reformulates the standard Transformer mechanism~\cite{vaswani2017attention} by representing input channels as tokens rather than temporal positions. This channel-oriented tokenization enables the model to directly capture inter-variable relationships across the multivariate sequence, an ability that proves especially effective for long-horizon forecasting where complex cross-channel dependencies must be modeled efficiently.

Building on this foundation, we replace the self-attention layers of the iTransformer with a QSANN module. The QSANN was originally proposed for text classification tasks~\cite{Li2024QSANN}, where it demonstrated the ability to capture semantic dependencies via VQCs. In this work, we reformulate the QSANN mechanism for multivariate time-series forecasting. The resulting quantum layers provide a quantum-native mechanism for encoding correlations and capturing higher-order interactions across variables. By integrating QSANN into the iTransformer backbone, the resulting iQTransformer constitutes a hybrid architecture that unites channel-wise tokenization with a quantum self-attention mechanism, specifically adapted to the challenges of multivariate time-series forecasting.

\subsection{iTransformer}
\label{subsec:itransformer}

\begin{figure}[t]
    \centering
    \includegraphics[width=1.0\linewidth]{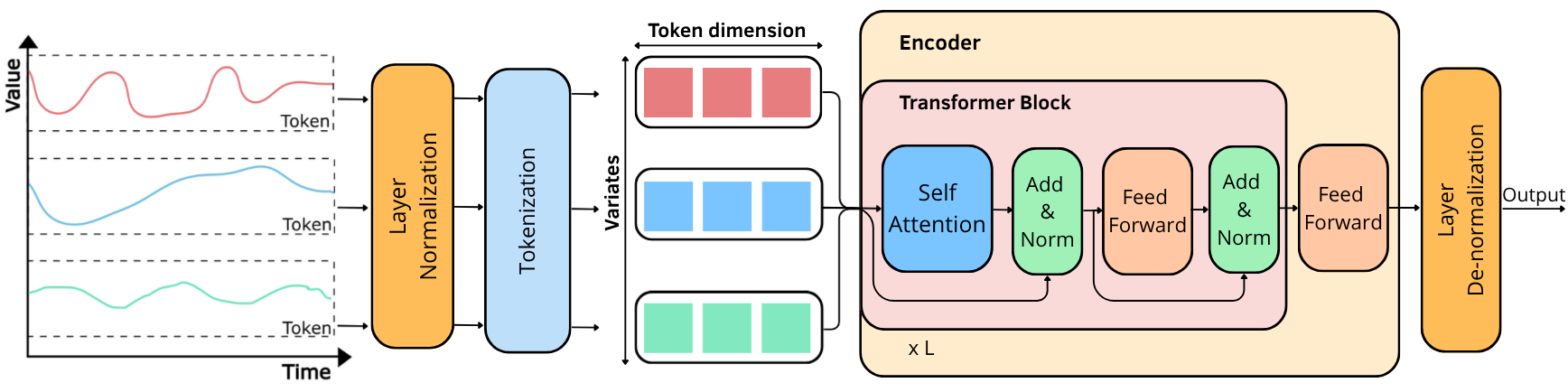}
    \caption{Schematic of the iTransformer/iQTransformer forecasting pipeline. Each variate is tokenized, processed through $L$ stacked Transformer blocks with self-attention across variates and feed-forward networks across time, and finally projected to the prediction horizon. In the iTransformer the Self-Attention block corresponds to the classical mechanism, whereas in the iQTransformer it is replaced by a Quantum Self-Attention Layer (QSAL).
    }
    \label{fig:itransformer_pipeline}
\end{figure}

Unlike the vanilla Transformer, which treats each time step as a token, the iTransformer~\cite{liu2023itransformer} encodes each variable (feature dimension) $\mathbf{X}_{:,c} \in \mathbb{R}^{T}$ as a token, effectively transposing the tokenization perspective to model inter-variable dependencies rather than temporal ones. In this formulation, the attention mechanism operates across variates to capture cross-variable dependencies, while temporal patterns are modeled by shared feed-forward networks applied independently to each token. This removes positional encodings and naturally represents multivariate time series. Formally, the forecasting pipeline is
\begin{subequations}\label{eq:itransformer_pipeline}
\begin{align}
\X'_{:,c} &= \mathrm{LN}(\X_{:,c}) \,, \\
h_c^{0} &= \phi_{\mathrm{token}}(\X'_{:,c}) \,, \\
H^{\ell+1} &= \mathrm{TrmBlock}(H^{\ell}) \,, \quad \ell=0,\ldots,L-1 \,, \\
\Y'_{:,c} &= \phi_{\mathrm{proj}}(h_c^{L}) \,, \\
\hat{\Y}_{:,c} &= \mathrm{LN}^{-1}(\Y'_{:,c}) \,,
\end{align}
\end{subequations}
where $H^{\ell} = [h^{\ell}_{1};\ldots;h^{\ell}_{C}] \in \mathbb{R}^{C\times D}$ is the matrix of $C$ variate tokens at layer $\ell$, and $\phi_{\mathrm{token}}:\mathbb{R}^{T}\to\mathbb{R}^{D}$ and $\phi_{\mathrm{proj}}:\mathbb{R}^{D}\to\mathbb{R}^{S}$ are MLPs acting along the temporal dimension, with $D$ the token dimension. $\mathrm{LN}$ denotes layer normalization that is applied independently to each variate token,
\begin{equation}
\mathrm{LN}(h_c) = \frac{h_c - \mu(h_c)}{\sqrt{\sigma^2(h_c) + \varepsilon}} \,, \quad c=1,\ldots,C \,,
\label{eq:LN}
  \end{equation}
where $\mu(h_c)$ and $\sigma^2(h_c)$ denote the mean and variance of the $T$-dimensional features, while $\mathrm{LN}^{-1}(\cdot)$ denotes the inverse normalization operator (denormalization). This reduces discrepancies across variates and improves stability during training \cite{nonstationarytransformerse}. The overall structure of the iTransformer is illustrated in Fig.~\ref{fig:itransformer_pipeline}.

Each Transformer block combines single-head self-attention across variates with a position-wise feed-forward network (FFN). With residual connections and post-normalization, the update is
\begin{align} 
\tilde{H}^{\ell} &= H^{\ell} + \mathrm{SelfAttn}(\mathrm{LN'}(H^{\ell})) \,, \label{eq:classic_att} \\
H^{\ell+1} &= \tilde{H}^{\ell} + \mathrm{FFN}(\mathrm{LN'}(\tilde{H}^{\ell})) \,.
\end{align}
Here $\mathrm{LN'}(\cdot)$ refers to the standard layer normalization introduced in Eq.~\eqref{eq:LN}, with the difference that it operates along the channel dimension $C$, while $\mathrm{LN}$ in Eq.~\eqref{eq:LN} is applied along the temporal dimension $T$.

Within the self-attention layer, given $H^{\ell}\in\mathbb{R}^{C\times D}$, the queries, keys, and values are obtained via linear projections
\begin{equation}
Q = H^{\ell} \mathbf{W}_Q \,, \quad K = H^{\ell} \mathbf{W}_K \,, \quad V = H^{\ell} \mathbf{W}_V \,,
\end{equation}
with $\mathbf{W}_Q, \mathbf{W}_K, \mathbf{W}_V \in \mathbb{R}^{D\times d_k}$. Here $d_k$ denotes the projected dimension.\footnote{Since iTransformer employs a single-head attention mechanism, in this case $d_k = D$, while in general multi-head attention one typically has $d_k = D/M$ for $M$ heads.} The attention weights are computed as
\begin{equation}
A = \mathrm{Softmax}\!\left(\frac{QK^{\top}}{\sqrt{d_k}}\right) \in \mathbb{R}^{C\times C} \,,
\end{equation}
and the attention output is
\begin{equation}
\mathrm{SelfAttn}(H^{\ell}) = AV \in \mathbb{R}^{C\times d_k} \,.
\end{equation}

The feed-forward network acts independently on each variate token and consists of two affine transformations with a ReLU activation in between given by
\begin{equation}
\mathrm{FFN}(h) = \mathbf{W}_2 \,\text{ReLU}(\mathbf{W}_1 h + b_1) + b_2 \,, \qquad h \in \mathbb{R}^D \,,
\end{equation}
where $\mathbf{W}_1,\mathbf{W}_2$ and $b_1,b_2$ are trainable parameters. 
The first affine layer expands the token from $\mathbb{R}^D$ to $\mathbb{R}^{D_\mathrm{ff}}$ (where $D_\mathrm{ff}$ is the internal dimension of the FFN), and the second layer projects it back to $\mathbb{R}^D$. After passing through the stack of $L$ Transformer blocks, the output tokens are projected to the forecasting horizon by $\phi_{\mathrm{proj}}$, and the predictions are finally rescaled to the original scale by applying the inverse normalization operator $\mathrm{LN}^{-1}(\cdot)$ introduced above.

Thus, the iTransformer captures multivariate correlations in an interpretable way, well-suited for time-series forecasting~\cite{liu2023itransformer}.

\subsection{Quantum Self-Attention Mechanism}
\label{subsec:qsann}

The Quantum Self-Attention Neural Network (QSANN)~\cite{Li2024QSANN} was originally proposed for text classification, where it was introduced as a quantum analogue of the self-attention mechanism by replacing the linear projections of queries, keys, and values with VQCs. Each quantum self-attention layer (QSAL) proceeds in three stages: (i) encoding the classical inputs into quantum states, (ii) applying distinct VQCs corresponding to queries, keys, and values, and (iii) computing quantum self-attention coefficients via Gaussian-projected measurements.

\begin{figure}[t] \centering \includegraphics[width=0.6\linewidth]{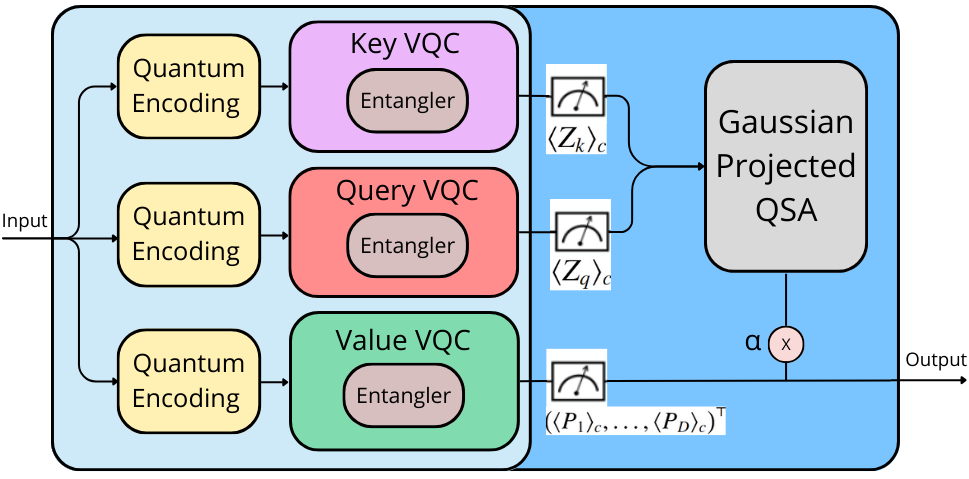} \caption{Schematic of a quantum self-attention layer (QSAL). Classical inputs are encoded into quantum states, processed by variational circuits representing query, key, and value, and then combined via Gaussian-projected self-attention coefficients.} \label{fig:qsann_layer} \end{figure} 

Given an input token $h_c^{\ell-1} \in \mathbb{R}^D$ at layer $\ell$, it is encoded into an $n$-qubit quantum state
\begin{equation}
|\psi_c\rangle = U_{\mathrm{enc}}(h_c^{\ell-1}) H^{\otimes n} |0^n\rangle \,, \quad c=1,\ldots,C \,,
\end{equation}
where $U_{\mathrm{enc}}$ is a quantum data-encoding ansatz. 

Three parameterized circuits $U_q(\theta_q)$, $U_k(\theta_k)$, and $U_v(\theta_v)$ represent the query, key, and value transformations. For each encoded state $|\psi_c\rangle$, the query and key associated with token $c$ are represented here as Pauli-$Z$ expectation values on a single designated qubit (here denoted as qubit 0),
\begin{equation}
\begin{aligned}
q_c = \langle Z_q \rangle_c = \langle \psi_c | U_q^\dagger(\theta_q) Z_0 U_q(\theta_q) | \psi_c \rangle \,, \\
k_c = \langle Z_k \rangle_c = \langle \psi_c | U_k^\dagger(\theta_k) Z_0 U_k(\theta_k) | \psi_c \rangle \,,    
\end{aligned}
\end{equation}
which are single real numbers in contrast to the $D$-dimensional query and key vectors used in the classical iTransformer. 
The value corresponding to token $c$ is represented by a $D$-dimensional vector of expectation values,
\begin{equation}
v_c = \big(\langle P_1 \rangle_c, \ldots, \langle P_D \rangle_c \big)^\top \,, \qquad 
\langle P_j \rangle_c = \langle \psi_c | U_v^\dagger(\theta_v) P_j U_v(\theta_v) | \psi_c \rangle \,,
\end{equation}
with $\{P_j\}$ a fixed set of Pauli observables.

After all circuits are evaluated for each token, the attention coefficients are computed by Gaussian projection of the query and key measurements,
\begin{equation}
\alpha_{c,c'} = \exp\!\big(-(\langle Z_q \rangle_c - \langle Z_k \rangle_{c'})^2\big) \,, \qquad
\tilde{\alpha}_{c,c'} = \frac{\alpha_{c,c'}}{\sum_{m=1}^C \alpha_{c,m}} \,,
\end{equation}
and the layer output is updated as
\begin{equation}
h_c^{\ell} = h_c^{\ell-1} + \sum_{c'=1}^C \tilde{\alpha}_{c,c'}\, v_{c'} \,.
\label{eq:qsann_update}
\end{equation}
This update is applied for every token $c=1,\ldots,C$. 
Eq.~\eqref{eq:qsann_update} is the quantum analogue of Eq.~\eqref{eq:classic_att}, replacing classical self-attention with its quantum counterpart. 
The overall structure of a QSAL is illustrated in Fig.~\ref{fig:qsann_layer}.

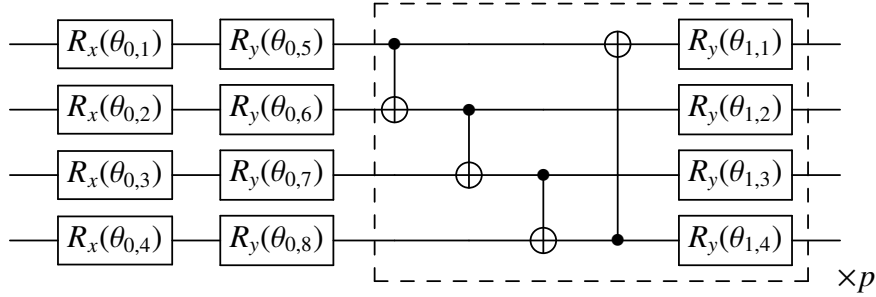
\begin{figure}[t] \centering \[\Qcircuit @C=1.5em @R=0.5em { & \gate{R_x(\theta_{0,1})} & \gate{R_y(\theta_{0,5})} &\ctrl{+1}& \qw & \qw &\targ & \gate{R_y(\theta_{1,1})}&\qw \\ & \gate{R_x(\theta_{0,2})} & \gate{R_y(\theta_{0,6})} & \targ & \ctrl{+1} &\qw&\qw & \gate{R_y(\theta_{1,2})} & \qw \\ & \gate{R_x(\theta_{0,3})} & \gate{R_y(\theta_{0,7})} & \qw &\targ& \ctrl{+1} &\qw& \gate{R_y(\theta_{1,3})} & \qw \\ & \gate{R_x(\theta_{0,4})} & \gate{R_y(\theta_{0,8})} & \qw&\qw &\targ & \ctrl{-3} & \gate{R_y(\theta_{1,4})} & \qw \gategroup{1}{4}{4}{8}{1.0em}{--} \\ &&&&&&&& \ \ \ \ \times p }\] \caption{Variational quantum ansatz used in QSANN. Layers of single-qubit rotations and entangling CNOT gates are repeated $p$ times to enhance expressivity.} \label{fig:qsann_ansatz} \end{figure}

QSANN employs the same ansatz for $U_{\mathrm{enc}}, U_q, U_k,$ and $U_v$. 
This ansatz, illustrated here in Fig.~\ref{fig:qsann_ansatz}, consists of layers of single-qubit $R_x$ and $R_y$ rotations followed by entangling CNOT gates, repeated $p$ times to enhance expressivity. 
When applied to $n$ qubits with depth $p$, the resulting encoding dimension is
\begin{equation}
D = n (p+2) \,,
\end{equation}
which corresponds to the number of independent classical parameters that can be embedded through $U_{\mathrm{enc}}$. 
In the case of the query, key, and value circuits $U_q$, $U_k$, and $U_v$, the same ansatz structure is adopted, 
and $D$ also represents the number of trainable variational parameters for each of these circuits. Thus, $D$ plays a dual role: as the feature dimension of the encoded tokens and as the parameter count per variational circuit.

For the value vectors $v_c$, the operators $\{P_j\}$ are taken from a fixed set of Pauli observables, and their choice determines the output dimension $D$. 
In the simplest setting with encoding depth $p=1$, one can select local measurements such as $\{X_i, Y_i, Z_i\}$ on each qubit $i=1,\ldots,n$, 
which yields $D = 3n$ observables in total. 

For deeper encodings ($p>1$), additional two-qubit observables (e.g., $Z_{12}$, $Z_{23}$, $\ldots$, where $Z_{ij} = Z_i \otimes Z_j$) can also be included to increase expressivity. 
Thus, the construction of $\{P_j\}$ provides flexibility in balancing the descriptive power of the value representation with the measurement cost. Further details on the design and implementation are given in~\cite{Li2024QSANN}.

\subsection{iQTransformer Model}
\label{subsec:iqtransformer}

The iQTransformer combines the inverted tokenization scheme of the iTransformer with the Quantum Self-Attention Neural Network (QSANN) mechanism described in Subsec.~\ref{subsec:qsann}. 
In this hybrid model, each channel $\X_{:,c}$ is encoded as a token, and the standard self-attention operator across channels is replaced by a quantum self-attention layer implemented via variational quantum circuits. 
This integration captures cross-channel dependencies in a quantum-enhanced space while preserving the efficient iTransformer structure.

Formally, the iQTransformer forecasting pipeline is defined as
\begin{subequations}\label{eq:iqtransformer_pipeline}
\begin{align}
\X'_{:,c} &= \mathrm{LN}(\X_{:,c}) \,, \\
h_c^{0} &= \phi_{\mathrm{token}}(\X'_{:,c}) \,, \\
H^{\ell+1} &= \mathrm{QTrmBlock}(H^{\ell}) \,, \quad \ell=0,\ldots,L-1 \,, \\
\Y'_{:,c} &= \phi_{\mathrm{proj}}(h_c^{L}) \,, \\
\hat{\Y}_{:,c} &= \mathrm{LN}^{-1}(\Y'_{:,c}) \,,
\end{align}
\end{subequations}
where the token encoding $\phi_{\mathrm{token}}$ and projection $\phi_{\mathrm{proj}}$ are defined as in Eq.~\eqref{eq:itransformer_pipeline}, 
and $\mathrm{QTrmBlock}(\cdot)$ denotes a Transformer block in which the self-attention operator is replaced by a Quantum Self-Attention Layer (QSAL).

Each quantum Transformer block updates the token representations as
\begin{align}
\tilde{H}^{\ell} &= H^{\ell} + \mathrm{QSAL}(\mathrm{LN'}(H^{\ell})) \,, \\
H^{\ell+1} &= \tilde{H}^{\ell} + \mathrm{FFN}(\mathrm{LN'}(\tilde{H}^{\ell})) \,,
\end{align}
so that, in analogy with Eq.~\eqref{eq:classic_att}, only the attention mechanism is replaced by its quantum counterpart $\mathrm{QSAL}$, while layer normalization, residual connections, and the feed-forward networks remain unchanged. 
Thus, the iQTransformer mirrors Fig.~\ref{fig:itransformer_pipeline}, with Self-Attention replaced by $\mathrm{QSAL}$.

\section{Experimental Setup} \label{sec:setup}

This section describes the datasets used in our experiments and the training procedure adopted for all models.  

\subsection{Datasets}
\label{subsec:datasets}

We evaluated our models on both synthetic and real-world datasets. 
For the synthetic setting, we used the Lorenz system, following the configuration described in~\cite{rivera20231d}. 
The Lorenz equations are defined as
\begin{align}
\dot{x} &= \sigma(y - x) \,, \\
\dot{y} &= -y - zx + \rho x \,, \\
\dot{z} &= -\beta z + xy \,,
\end{align}
where $(x,y,z)$ are the state variables and $\sigma, \rho, \beta$ are the system parameters. 
The system is solved by the Euler method with parameters $\sigma=10$, $\rho=28$, $\beta=8/3$ and initial point $(0,-0.01,9)$, generating a 1000-point trajectory as the synthetic benchmark.

For the real-world evaluation, we used a dataset from the energy production sector, namely the ITER dataset. This dataset covers a four-month period, from January to April 2024, and consists of operational and meteorological measurements from a wind turbine (AERO01) located in the southeast area of Tenerife, Canary Islands (Spain). Data were collected at a fixed sampling frequency of 15 minutes. Although the original dataset contained a timestamp, this attribute was removed in order to treat the series as equidistant temporal sequences, thus avoiding bias from irregularities in time indexing. The preprocessing applied to this dataset is described below.

All variables were normalized to ensure comparability and facilitate model convergence, with the exception of the operating state of the turbine variable. In this case, a value of 0 indicates normal operation, while values above 100 correspond to turbine shutdown events. Records associated with shutdown states were excluded from the dataset to avoid contaminating the training distribution.  

The power output variable was normalized by dividing by the rated capacity of the turbine. The curtailment setpoint was adjusted so that its maximum value coincided with the nominal capacity of the wind farm. Wind speed was restricted to physically valid values below or equal to 25~m/s, ensuring the elimination of spurious or sensor-related errors.

Regarding missing values, interpolation was applied only when gaps were limited in duration; segments with more than eight consecutive missing samples were excluded to prevent the introduction of artifacts in the temporal dynamics.  

After all preprocessing steps, the final dataset comprised seven channels: total energy demand, renewable energy production in Tenerife, percentage of renewable energy, normalized power, wind speed, wind direction (with the 0°/360° boundary properly normalized), and the curtailment setpoint (i.e., the limit imposed on the turbine's power output). The latter constitutes the target variable for prediction in this use case. 

\subsection{Training Procedure}
\label{subsec:training}

All models were trained in a hybrid framework using \texttt{PennyLane}~\cite{bergholm2018pennylane} for simulation and \texttt{PyTorch}~\cite{paszke2019pytorch} for gradient-based optimization. The \texttt{default.qubit} statevector backend was employed for simulating quantum operations. Optimization was performed with the Adam optimizer, using a learning rate of 
$5 \times 10^{-4}$ and mean-squared error (MSE) as the loss function. Models were trained for 50 epochs and 250 epochs for Lorenz and ITER datasets respectivelly with random initialization; each experiment was repeated with 10 seeds and results averaged. Datasets were split 75\%/25\% for training/validation.

Experiments were conducted under two forecasting regimes: short-term (ST) and long-term (LT). For the Lorenz dataset, in the ST setting a training window of $T=5$ past points was used to predict a single future step ($S=1$), while in the LT case $T=5$ past points were used to predict $S=5$ steps ahead across all channels. A batch size of $128$ was employed in this dataset. 

For the ITER dataset, we adopted a forecasting setup closer to a realistic application in wind turbine energy production. Here, the seven available channels were used to predict a single target channel corresponding to the curtailment setpoint. In the ST setting, $T=5$ past points were used to predict the next step. In the LT setting, a much wider input horizon of $T=336$ observations (corresponding to $3.5$ days of data) was employed to forecast $S=24$ future steps (equivalent to $6$ hours) on the target channel. A batch size of $1024$ was used in this case.

In addition to the transformer-based models (iTransformer \& iQTransformer) introduced in Sec.~\ref{sec:base}, and the multichannel forecasting strategies of Sec.~\ref{sec:approaches}, we included baselines for a fair comparison with other state-of-the-art approaches. Specifically, we considered a classical one-dimensional CNN model inspired by Ref.~\cite{rivera20231d} and a quantum recurrent architecture based on quantum gated recurrent units (QGRU) following Ref.~\cite{chen2024quantum}. The QGRU baseline followed an encoder-decoder scheme: the encoder maps inputs to quantum space, QGRU models temporal dependencies, and the decoder projects outputs back to classical space.

For all VQC-based architectures introduced in Sec.~\ref{sec:approaches}, a fixed number of layers $p=24$ was used. In the encoder–decoder VQC model, the circuit was implemented with $n=8$ qubits. For the transformer-based models, two configurations were adopted depending on the dataset. In the Lorenz case, we used $L=2$ transformer blocks with embedding dimension $D=9$, feed-forward dimension $D_{\mathrm{ff}}=12$, and $n=3$ qubits, with encoding and variational circuit depths $p_{\mathrm{enc}}=1$ and $p_{\mathrm{vqc}}=3$, respectively. For the ITER dataset, the configuration was set to $D=16$, $D_{\mathrm{ff}}=8$, $n=4$, $p_{\mathrm{enc}}=2$, and $p_{\mathrm{vqc}}=3$. Although both iTransformer and iQTransformer share the same hyperparameters, their trainable parameter counts differ due to the use of quantum attention layers. The iTransformer comprises 1877 and 1929 parameters for the Lorenz short- and long-term settings, and 4441 and 11445 for the ITER dataset. In contrast, the iQTransformer contains 719 and 771 parameters for the Lorenz cases, and 1357 and 5295 for the ITER runs: less than half of those of the classical model.

\section{Results} \label{sec:results}

\begin{table}[t]
    \centering
    \begin{tabular}{|l|cc|cc|cc|}
    \hline
        \multirow{2}{*}{Model} & \multicolumn{2}{c|}{MAPE} & \multicolumn{2}{c|}{MAE} & \multicolumn{2}{c|}{RMSE} \\
        \cline{2-7}
                               & Lorenz & ITER & Lorenz & ITER  & Lorenz & ITER \\
    \hline
    \multicolumn{7}{|c|}{\textbf{Short-Term Forecasting}} \\
    \hline
    VQC (indep.) & 0.0353 & 0.1490   & 0.0173 & 0.0905   & 0.0281 & 0.1341 \\
    VQC + MLP         & 0.305 & 0.1037   &0.139 & 0.0522  &   0.184 & 0.0963 \\
    DE. (obs.)  & 0.117 & n/a   & 0.0558 & n/a   & 0.0982 & n/a \\
    DE. (qubits)  & 0.116 & n/a   & 0.0563 & n/a   & 0.0985 & n/a \\
    Enc.–VQC–Dec.     & 0.454 & 0.3800  & 0.210 & 0.3435   & 0.259 & 0.3574\\
    Data re-upload. & 0.0388 & 0.0518   & 0.0192 & 0.0353   & 0.0308 & 0.0534 \\
    QGRU              & 0.432 & 0.1852   & 0.195 &  0.1205   & 0.243 & 0.1577 \\
    1D CNN            & 0.204 & 0.0798   & 0.082 & 0.0434   &  0.101 & 0.0741 \\
    iTransformer      & $\mathbf{0.0081}$ & 0.0154  & $\mathbf{0.0039}$ &  $\mathbf{0.0069}$  & $\mathbf{0.0064}$ & 0.0354 \\
    iQTransformer     & 0.0086 & $\mathbf{0.0152}$   & 0.0041 & $0.0071$   & 0.0067 & $\mathbf{0.0351}$ \\
    \hline
    \multicolumn{7}{|c|}{\textbf{Long-Term Forecasting}} \\
    \hline
    VQC + MLP         & 0.235 & n/a   & 0.0976 & n/a   & 0.127 & n/a \\
    Enc.–VQC–Dec.      & 0.283 & 0.0947   & 0.114 & 0.0477  & 0.141 & 0.1076\\
    1D CNN            & 0.235 & 0.1191   & 0.0954 &  0.0515   & 0.122 & 0.1241 \\
    iTransformer      & 0.0498 & 0.0874  &  0.0234 & 0.0352  & 0.0371 & 0.1050 \\
    iQTransformer     & $\mathbf{0.0490}$ & $\mathbf{0.0849}$ & $\mathbf{0.0230}$ & $\mathbf{0.0340}$ & $\mathbf{0.0364}$ & $\mathbf{0.1019}$ \\
    \hline
    \end{tabular}
    \caption{Short-term and long-term forecasting performance on the validation set. Values 
    are averaged over the final 10 training epochs and over 10 random initializations of each model.
    }
    \label{tab:results}
\end{table}

This section presents the forecasting performance of quantum and classical models for both short-term (ST) and long-term (LT) prediction on the Lorenz and ITER datasets, under the setup described in Sec.~\ref{sec:setup}. The evaluated models include independent VQCs, VQC+MLP, dense embedding (Lorenz only), encoder–decoder VQC, data re-uploading, QGRU, 1D CNN, iTransformer, and iQTransformer. Some architectures were only evaluated in the ST scenario due to design constraints that prevent their straightforward extension to multi-step forecasting.

Table~\ref{tab:results} summarizes the forecasting performance of all evaluated models, measured using mean absolute percentage error (MAPE), mean absolute error (MAE), and root mean squared error (RMSE). Values correspond to validation results averaged over the final 10 training epochs.

\begin{figure}[t]
    \centering
    \includegraphics[width=1.0\textwidth]{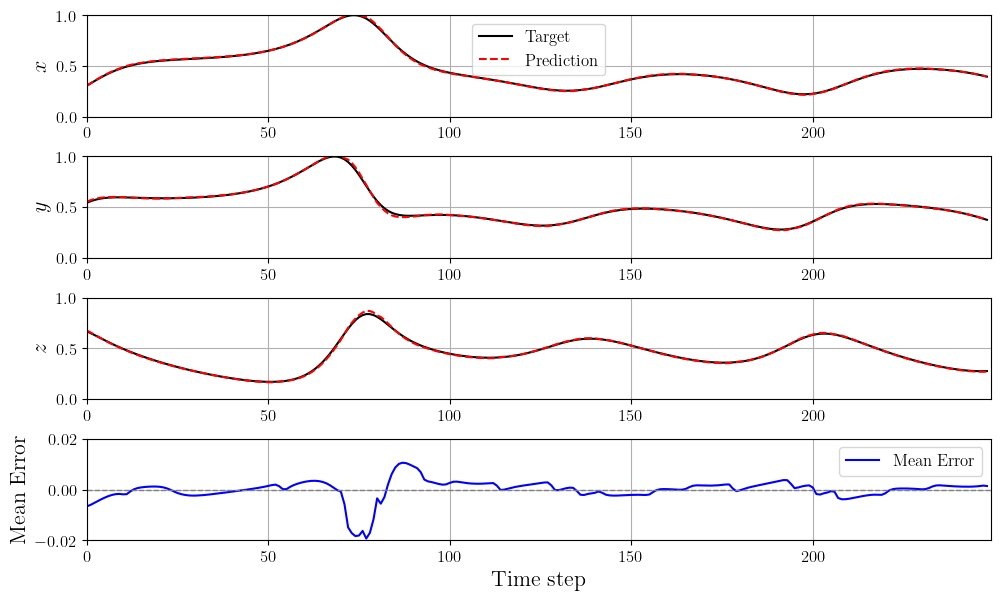}
    \caption{Short-term reconstruction of the Lorenz dataset using the best performing iQTransformer model execution. Black curves indicate the ground-truth series, while \textcolor{red}{red} curves denote the reconstructed predictions obtained from the previous five ground-truth points. The bottom panel shows the reconstruction error averaged across all channels, displayed in \textcolor{blue}{blue}.}
    \label{fig:st_reconstruction}
\end{figure}

For the ST horizon ($S=1$), the transformer-based architectures achieved the lowest error metrics across both datasets. The iTransformer and iQTransformer outperform all quantum and classical baselines, with the iTransformer slightly leading on the Lorenz dataset and the iQTransformer showing the best performance on the ITER dataset. Notably, the iQTransformer attains this accuracy with fewer than half the trainable parameters of the iTransformer, highlighting its efficiency in representation and optimization. Among purely quantum circuits, the data re-uploading model exhibited strong and consistent performance across runs, while other models obtained comparatively higher errors. This is attributed not only to architectural limitations, but to the inherently faster convergence of transformer-based architectures, which reach near-optimal performance with fewer epochs thanks to their strong ability to capture temporal dependencies and long-range correlations.

Fig.~\ref{fig:st_reconstruction} illustrates the reconstruction of the Lorenz dataset obtained with the iQTransformer model on the validation set. The model achieves high-quality predictions across all three channels, closely following the ground truth trajectories over the full sequence. The bottom panel shows the reconstruction error, which remains of order $\lesssim 10^{-2}$ in absolute value, substantially smaller than the normalized series values of magnitude $\sim 1$. This indicates that the iQTransformer is able to capture the underlying dynamics with high fidelity in the short-term regime.

\begin{figure}[t]
    \centering
    \includegraphics[width=0.85\linewidth]{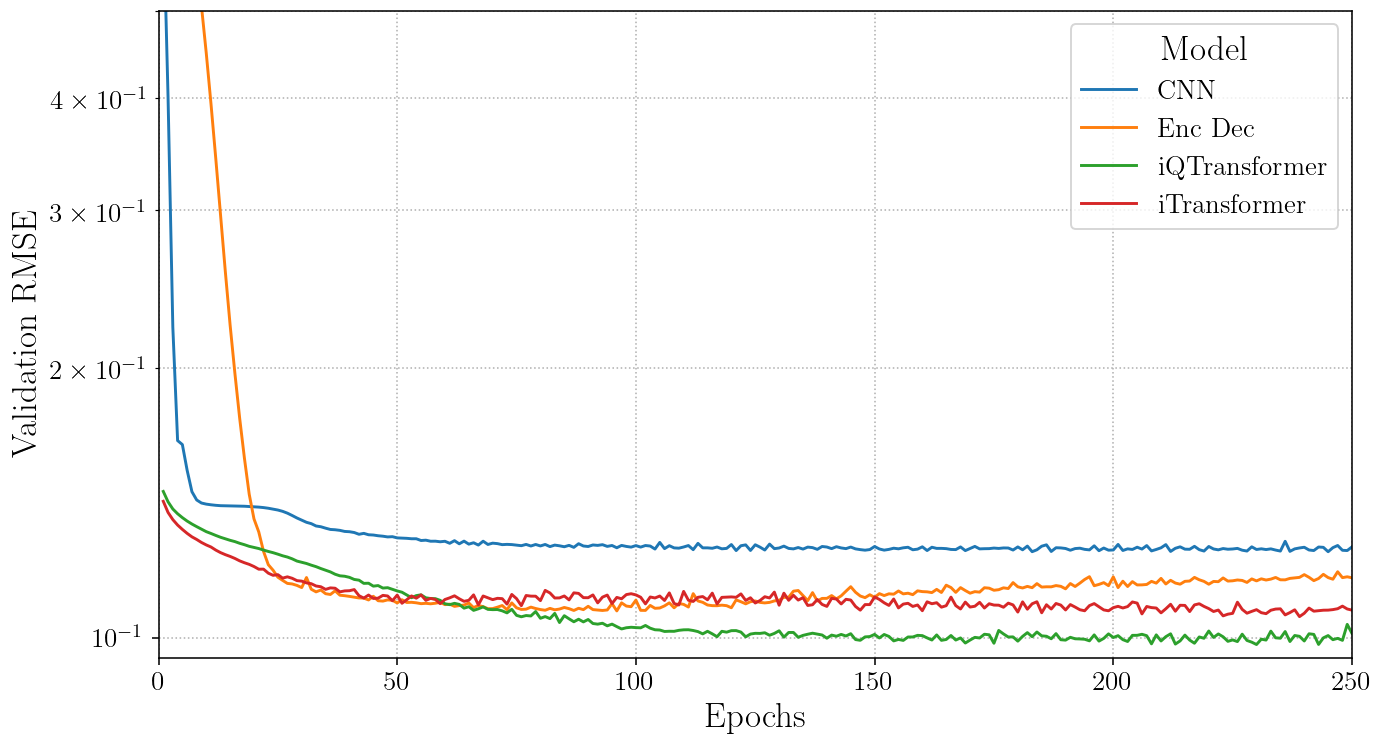}
    \caption{Validation RMSE curves vs. training epochs for the CNN, Encoder-Decoder (Enc Dec), iQTransformer and iTransformer models on the ITER dataset for the LT Forecasting.}
    \label{fig:convergence}
\end{figure}

\begin{figure}[t]
    \centering
    \includegraphics[width=0.80\textwidth]{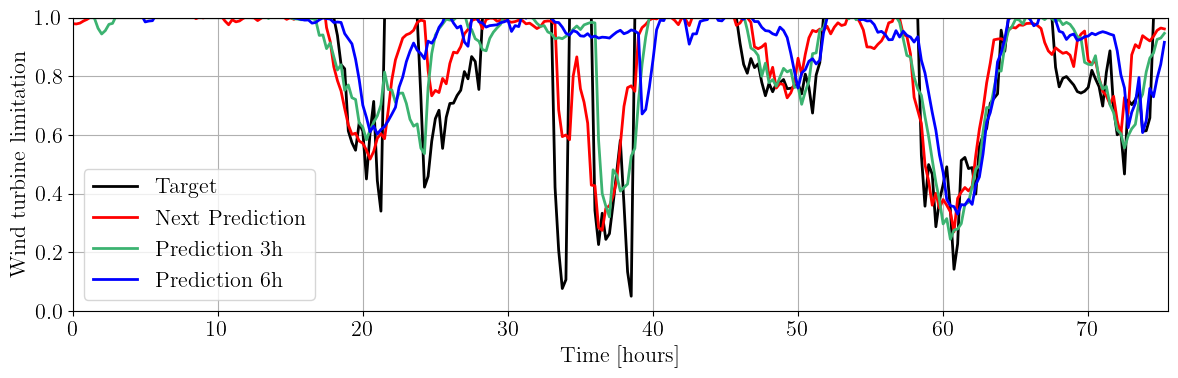}
    \caption{Long-term forecasting on the ITER validation set using the best performing iQTransformer model execution. The black curve indicates the ground-truth target channel associated with turbine power limitation. Colored curves show the predictions at different horizons: \textcolor{red}{red} for the immediate next-step, \textcolor{mediumseagreen}{green} for 3-hour forecasts, and \textcolor{blue}{blue} for 6-hour forecasts.}
    \label{fig:lt_reconstruction}
\end{figure}

For the long-term prediction regime ($S>1$), only models with explicit temporal memory or hierarchical representation maintained stable performance. The iQTransformer achieved the best results on both Lorenz and ITER datasets, closely followed by the iTransformer. Classical baselines such as the 1D CNN displayed stronger degradation with increasing horizon length.  Fig.~\ref{fig:convergence} shows the RMSE on the validation set versus the training epochs, illustrating the fast convergence of transformer models.

Fig.~\ref{fig:lt_reconstruction} illustrates long-term forecasting results on the ITER validation set using the iQTransformer model. The black curve denotes the ground-truth target channel, while the red, green, and blue curves show the predictions at horizons of one step, 3h (12 steps), and 6h (24 steps), respectively. The model reproduces the main fluctuations of the target signal, correctly anticipating most limitation peaks. In particular, the pronounced event around $60$h is predicted with high precision at a 6h horizon, both in timing and magnitude. Other peaks, such as the event near $35$h, are accurately forecasted at 3h but not fully captured at 6h. Overall, the iQTransformer generalizes across extended horizons and performs well in realistic use cases, such as wind turbine limitation forecasting.

\section{Conclusions} \label{sec:conclusions}

In this work, we investigated quantum machine learning (QML) architectures for multivariate time-series forecasting, extending their applicability beyond the predominantly univariate settings explored in prior research. We benchmarked several VQC-based approaches, including independent-channel designs, dense embeddings, hybrid encoder-decoder formulations, and data re-uploading models, as well as a recurrent QGRU baseline. To provide a fair comparison, these quantum models were evaluated alongside classical architectures such as 1D CNNs and the iTransformer.

Building upon these baselines, we proposed the iQTransformer, a novel hybrid model that integrates a Quantum Self-Attention Neural Network (QSANN) into the iTransformer framework. This design leverages channel-wise embeddings and quantum-native self-attention to capture cross-variable dependencies in a more expressive feature space. Across both synthetic (Lorenz) and real-world (ITER) datasets, the iQTransformer achieved superior accuracy compared to classical baselines, while retaining efficiency advantages in terms of training dynamics: fast convergence dynamics and fewer number of trainable parameters.

Our results highlight two key takeaways. First, quantum-enhanced models may compete with strong classical baselines in practical forecasting scenarios, benefiting from efficient training and compact parameterization. Second, hybrid quantum-classical designs offer a promising balance: quantum circuits capture higher-order correlations, while classical components provide scalability and stability for long-horizon forecasting.

Overall, this study demonstrates that QML, and in particular the proposed iQTransformer, represents a promising step towards efficient, scalable, and interpretable multivariate forecasting with quantum-enhanced models.

\section*{Acknowledgments}
This work was partially supported by Grant PID2023-146520OB-C22, funded by MCIU/AEI/10.13039/501100011033, by Grant IDE/2024/000734, funded by the Principality of Asturias, and by the QUANTUM ENIA project call – Quantum Spain project, funded by the Spanish Ministry for Digital Transformation and of Civil Service and the European Union (NextGenerationEU, Digital Spain 2026 Agenda); and by ARQUIMEA Research Center and Horizon Europe, Teaming for Excellence, under grant agreement No 101059999, project QCircle. The work of Jesús Bonilla was partially supported by the Spanish Ministry of Science, Innovation and Universities through the Torres Quevedo grant PTQ2023-013228. The work of Jorge Ballesteros is financially supported by Instituto Tecnológico y de Energías Renovables (ITER) and Cabildo de Tenerife.




\nocite{*}
\bibliographystyle{elsarticle-num-names} 
\bibliography{elsarticle}

\begin{thebibliography}{17}
\expandafter\ifx\csname natexlab\endcsname\relax\def\natexlab#1{#1}\fi
\providecommand{\url}[1]{\texttt{#1}}
\providecommand{\href}[2]{#2}
\providecommand{\path}[1]{#1}
\providecommand{\DOIprefix}{doi:}
\providecommand{\ArXivprefix}{arXiv:}
\providecommand{\URLprefix}{URL: }
\providecommand{\Pubmedprefix}{pmid:}
\providecommand{\doi}[1]{\href{http://dx.doi.org/#1}{\path{#1}}}
\providecommand{\Pubmed}[1]{\href{pmid:#1}{\path{#1}}}
\providecommand{\bibinfo}[2]{#2}
\ifx\xfnm\relax \def\xfnm[#1]{\unskip,\space#1}\fi
\bibitem[{Schuld and Petruccione(2021)}]{schuld2021machine}
\bibinfo{author}{M.~Schuld}, \bibinfo{author}{F.~Petruccione}, \bibinfo{title}{Machine learning with quantum computers}, \bibinfo{publisher}{Springer}, \bibinfo{year}{2021}.
\bibitem[{Biamonte et~al.(2017)Biamonte, Wittek, Pancotti, Rebentrost, Wiebe, and Lloyd}]{biamonte2017quantum}
\bibinfo{author}{J.~Biamonte}, \bibinfo{author}{P.~Wittek}, \bibinfo{author}{N.~Pancotti}, \bibinfo{author}{P.~Rebentrost}, \bibinfo{author}{N.~Wiebe}, \bibinfo{author}{S.~Lloyd},
\newblock \bibinfo{title}{Quantum machine learning},
\newblock \bibinfo{journal}{Nature} \bibinfo{volume}{549} (\bibinfo{year}{2017}) \bibinfo{pages}{195--202}.
\bibitem[{Bergholm et~al.(2018)Bergholm, Izaac, Schuld, Gogolin, Ahmed, Ajith, Alam, Alonso-Linaje, AkashNarayanan, Asadi et~al.}]{bergholm2018pennylane}
\bibinfo{author}{V.~Bergholm}, \bibinfo{author}{J.~Izaac}, \bibinfo{author}{M.~Schuld}, \bibinfo{author}{C.~Gogolin}, \bibinfo{author}{S.~Ahmed}, \bibinfo{author}{V.~Ajith}, \bibinfo{author}{M.~S. Alam}, \bibinfo{author}{G.~Alonso-Linaje}, \bibinfo{author}{B.~AkashNarayanan}, \bibinfo{author}{A.~Asadi}, et~al.,
\newblock \bibinfo{title}{Pennylane: Automatic differentiation of hybrid quantum-classical computations},
\newblock \bibinfo{journal}{arXiv preprint arXiv:1811.04968}  (\bibinfo{year}{2018}).
\bibitem[{Fellner et~al.(2025)Fellner, Kreplin, Tovey, and Holm}]{fellner2025quantum}
\bibinfo{author}{T.~Fellner}, \bibinfo{author}{D.~Kreplin}, \bibinfo{author}{S.~Tovey}, \bibinfo{author}{C.~Holm},
\newblock \bibinfo{title}{Quantum vs. classical: A comprehensive benchmark study for predicting time series with variational quantum machine learning},
\newblock \bibinfo{journal}{arXiv preprint arXiv:2504.12416}  (\bibinfo{year}{2025}).
\bibitem[{Zhou et~al.(2021)Zhou, Zhang, Peng, Zhang, Li, Xiong, and Zhang}]{zhou2021informer}
\bibinfo{author}{H.~Zhou}, \bibinfo{author}{S.~Zhang}, \bibinfo{author}{J.~Peng}, \bibinfo{author}{S.~Zhang}, \bibinfo{author}{J.~Li}, \bibinfo{author}{H.~Xiong}, \bibinfo{author}{W.~Zhang},
\newblock \bibinfo{title}{Informer: Beyond efficient transformer for long sequence time-series forecasting},
\newblock \bibinfo{journal}{Proceedings of the AAAI Conference on Artificial Intelligence} \bibinfo{volume}{35} (\bibinfo{year}{2021}) \bibinfo{pages}{11106--11115}. \URLprefix \url{https://ojs.aaai.org/index.php/AAAI/article/view/17325}. \DOIprefix\doi{10.1609/aaai.v35i12.17325}.
\bibitem[{Liu et~al.(2023)Liu, Hu, Zhang, Wu, Wang, Ma, and Long}]{liu2023itransformer}
\bibinfo{author}{Y.~Liu}, \bibinfo{author}{T.~Hu}, \bibinfo{author}{H.~Zhang}, \bibinfo{author}{H.~Wu}, \bibinfo{author}{S.~Wang}, \bibinfo{author}{L.~Ma}, \bibinfo{author}{M.~Long},
\newblock \bibinfo{title}{itransformer: Inverted transformers are effective for time series forecasting},
\newblock \bibinfo{journal}{arXiv preprint arXiv:2310.06625}  (\bibinfo{year}{2023}).
\bibitem[{Lim and Zohren(2021)}]{lim2021survey}
\bibinfo{author}{B.~Lim}, \bibinfo{author}{S.~Zohren},
\newblock \bibinfo{title}{Time-series forecasting with deep learning: a survey},
\newblock \bibinfo{journal}{Philosophical Transactions of the Royal Society A} \bibinfo{volume}{379} (\bibinfo{year}{2021}) \bibinfo{pages}{20200209}.
\bibitem[{Hochreiter and Schmidhuber(1997)}]{hochreiter1997lstm}
\bibinfo{author}{S.~Hochreiter}, \bibinfo{author}{J.~Schmidhuber},
\newblock \bibinfo{title}{Long short-term memory},
\newblock \bibinfo{journal}{Neural computation} \bibinfo{volume}{9} (\bibinfo{year}{1997}) \bibinfo{pages}{1735--1780}.
\bibitem[{Vaswani and et~al.(2017)}]{vaswani2017attention}
\bibinfo{author}{A.~Vaswani}, \bibinfo{author}{et~al.},
\newblock \bibinfo{title}{Attention is all you need},
\newblock \bibinfo{journal}{Advances in neural information processing systems} \bibinfo{volume}{30} (\bibinfo{year}{2017}).
\bibitem[{Liu et~al.(2023)Liu, Wu, Wang, and Long}]{nonstationarytransformerse}
\bibinfo{author}{Y.~Liu}, \bibinfo{author}{H.~Wu}, \bibinfo{author}{J.~Wang}, \bibinfo{author}{M.~Long},
\newblock \bibinfo{title}{Non-stationary transformers: Exploring the stationarity in time series forecasting}  (\bibinfo{year}{2023}). \URLprefix \url{https://arxiv.org/abs/2205.14415}. \href{http://arxiv.org/abs/2205.14415}{{\tt arXiv:2205.14415}}.
\bibitem[{Rivera-Ruiz et~al.(2023)Rivera-Ruiz, Ju{\'a}rez-Osorio, Mendez-Vazquez, L{\'o}pez-Romero, and Rodriguez-Tello}]{rivera20231d}
\bibinfo{author}{M.~A. Rivera-Ruiz}, \bibinfo{author}{S.~L. Ju{\'a}rez-Osorio}, \bibinfo{author}{A.~Mendez-Vazquez}, \bibinfo{author}{J.~M. L{\'o}pez-Romero}, \bibinfo{author}{E.~Rodriguez-Tello},
\newblock \bibinfo{title}{1d quantum convolutional neural network for time series forecasting and classification},
\newblock in: \bibinfo{booktitle}{Mexican International Conference on Artificial Intelligence}, \bibinfo{organization}{Springer}, \bibinfo{year}{2023}, pp. \bibinfo{pages}{17--35}.
\bibitem[{Cong et~al.(2019)Cong, Choi, and Lukin}]{cong2019quantum}
\bibinfo{author}{I.~Cong}, \bibinfo{author}{S.~Choi}, \bibinfo{author}{M.~D. Lukin},
\newblock \bibinfo{title}{Quantum convolutional neural networks},
\newblock \bibinfo{journal}{Nature Physics} \bibinfo{volume}{15} (\bibinfo{year}{2019}) \bibinfo{pages}{1273--1278}.
\bibitem[{Hur et~al.(2022)Hur, Kim, and Park}]{hur2022quantum}
\bibinfo{author}{T.~Hur}, \bibinfo{author}{L.~Kim}, \bibinfo{author}{D.~K. Park},
\newblock \bibinfo{title}{Quantum convolutional neural network for classical data classification},
\newblock \bibinfo{journal}{Quantum Machine Intelligence} \bibinfo{volume}{4} (\bibinfo{year}{2022}) \bibinfo{pages}{3}.
\bibitem[{P{\'e}rez-Salinas et~al.(2020)P{\'e}rez-Salinas, Cervera-Lierta, Gil-Fuster, and Latorre}]{perez2020data}
\bibinfo{author}{A.~P{\'e}rez-Salinas}, \bibinfo{author}{A.~Cervera-Lierta}, \bibinfo{author}{E.~Gil-Fuster}, \bibinfo{author}{J.~I. Latorre},
\newblock \bibinfo{title}{Data re-uploading for a universal quantum classifier},
\newblock \bibinfo{journal}{Quantum} \bibinfo{volume}{4} (\bibinfo{year}{2020}) \bibinfo{pages}{226}.
\bibitem[{Chen and Khaliq(2024)}]{chen2024quantum}
\bibinfo{author}{Y.~Chen}, \bibinfo{author}{A.~Khaliq},
\newblock \bibinfo{title}{Quantum recurrent neural networks: Predicting the dynamics of oscillatory and chaotic systems},
\newblock \bibinfo{journal}{Algorithms} \bibinfo{volume}{17} (\bibinfo{year}{2024}) \bibinfo{pages}{163}.
\bibitem[{Li et~al.(2024)Li, Zhao, and Wang}]{Li2024QSANN}
\bibinfo{author}{G.~Li}, \bibinfo{author}{X.~Zhao}, \bibinfo{author}{X.~Wang},
\newblock \bibinfo{title}{Quantum self-attention neural networks for text classification},
\newblock \bibinfo{journal}{Science China Information Sciences} \bibinfo{volume}{67} (\bibinfo{year}{2024}) \bibinfo{pages}{142501}. \URLprefix \url{https://doi.org/10.1007/s11432-23-3879-7}. \DOIprefix\doi{10.1007/s11432-023-3879-7}.
\bibitem[{Paszke et~al.(2019)Paszke, Gross, Massa, Lerer, Bradbury, Chanan, Killeen, Lin, Gimelshein, Antiga, Desmaison, Kopf, Yang, DeVito, Raison, Tejani, Chilamkurthy, Steiner, Fang, Bai, and Chintala}]{paszke2019pytorch}
\bibinfo{author}{A.~Paszke}, \bibinfo{author}{S.~Gross}, \bibinfo{author}{F.~Massa}, \bibinfo{author}{A.~Lerer}, \bibinfo{author}{J.~Bradbury}, \bibinfo{author}{G.~Chanan}, \bibinfo{author}{T.~Killeen}, \bibinfo{author}{Z.~Lin}, \bibinfo{author}{N.~Gimelshein}, \bibinfo{author}{L.~Antiga}, \bibinfo{author}{A.~Desmaison}, \bibinfo{author}{A.~Kopf}, \bibinfo{author}{E.~Yang}, \bibinfo{author}{Z.~DeVito}, \bibinfo{author}{M.~Raison}, \bibinfo{author}{A.~Tejani}, \bibinfo{author}{S.~Chilamkurthy}, \bibinfo{author}{B.~Steiner}, \bibinfo{author}{L.~Fang}, \bibinfo{author}{J.~Bai}, \bibinfo{author}{S.~Chintala},
\newblock \bibinfo{title}{Pytorch: An imperative style, high-performance deep learning library},
\newblock in: \bibinfo{booktitle}{Advances in Neural Information Processing Systems 32}, \bibinfo{year}{2019}, pp. \bibinfo{pages}{8024--8035}.

\end{thebibliography}

\end{document}